\documentclass{ws-ijmpd}

\begin{document}

\markboth{Spaans}
{The Role of Information in Gravity}

\catchline{}{}{}{}{}
%
%%%%%%%%%%%%%%%%%%%%%%%%%%%%%%%%%%%%%%%%%%%%%%%%%%%%%%%%%%%%%%%%%%%%

\title{The Role of Information in Gravity}

\author{M.\ Spaans}

\address{Kapteyn Institute, University of Groningen\\ P.O. Box 800, 9700 AV Groningen, Netherlands\\ spaans@astro.rug.nl}

%\author{SECOND AUTHOR}

%\address{Group, Laboratory, Address\\
%City, State ZIP/Zone, Country\\
%second\_author@group.com}

\maketitle

%\begin{history}
%\received{Day Month Year}
%\revised{Day Month Year}
%\comby{Managing Editor}
%\end{history}

\begin{abstract}
It is argued that particle-specific information on energy-momentum adjusts the
strength of gravity.
This form of gravity has no free parameters, preserves Einstein gravity locally
and predicts 6 times stronger accelerations on galaxy scales.
\end{abstract}

\section{Information and Gravity}

The crucial role played by information in quantum mechanics has been shown by
interference experiments between beams of light and atoms. Remarkably, a single
bit of information, attached as an identifier to one of the two slits in a
double slit experiment, suffices to destroy an interference pattern\cite{1}. In
this paper, the focus lies on the relevance of information for
gravity\cite{2,3}, and the dependence of Einstein gravity on information is
explored with a few simple arguments.

Consider first Einstein's famous elevator/rocket thought experiment, where one
compares an observer in a suspended elevator on Earth to an observer that is
accelerated by a rocket through empty space.
The operator of the rocket that is responsible for the equivalent acceleration
of the moving observer needs to be informed about the strength of the
gravitational field that pertains to the suspended observer, if he is to mimick
the proper experience.
Without this information the rocket operator cannot realize the thought
experiment as an actual physical event.

A priori, it is unclear what significance can be assigned to this assertion.
Nevertheless, it appears that information on a gravitational system, in some
form, is necessary to define a physical situation that mimicks gravity as an
acceleration field. This argument merely seeks to enlarge the equivalence
principle (EP, in any of its formulations 'weak', 'Einstein' or 'strong') with
a sense of constructability and does not affect the benefits of the EP in any
way, since the latter derives from the local indistinguishability of physical
experiences.

Put differently, every observer can free-fall through space-time under the
influence of gravity. This suggests that information is available to tell
observers how to move.
Indeed, mass curves space-time and curvature tells mass how to move\cite{2}.
However, this still requires information to characterize free-falling observers
physically.

Therefore, one is lead to the tempting position that the law of gravity is not
imposed on individual observers as an exterior rule, but that it is physically
imposed on them by the information that is actually necessary to realize
specific gravitational dynamics. This is referred to as the information
postulate (IP).
Hence, according to the IP, necessary information just as much implements
gravity is it expresses gravity: ``What one needs to know determines what one
experiences''.
The required information that is pertinent here, is the one on the
energy-momentum of particles since energy-momentum gravitates and matter is
undeniably made from particles.

This is a good point to be more precise about information.
For $n$ the, not necessarily constant, number of particles in some system, and
$w$ counting particle properties like space-time differential intervals, mass,
charge and spin, an obvious measure of information is $I\equiv nw+x$, with $x$
representing system wide properties.
One may think of a set of numbers that carries information pertinent to
individual particles and the system at large. Crucially, it only matters that a
particle has a mass, not what the magnitude of that mass is. Also, quantities
like energy and angular momentum are derivable from mass, and space and time
intervals or vice versa.

Note then that a free-falling observer, locally impervious to the force of
gravity, requires only non-gravitational information to be described himself,
i.e., using information pertaining to non-gravitational interactions.
Of course, gravitational information is needed for the energy-momentum induced
curvature of the global space-time that an observer occupies.
Hence, under the IP, it is the insufficiency of non-gravitational information
to define particle dynamics that imposes gravity.

To formalize and quantify this, one can introduce non-gravitational information
on energy-momentum, $N$, that describes particle interactions in the absence of
gravity, and gravitational information, $N'$, that defines the gravitational
interactions exerted by particles.
Interestingly, Einstein gravity is locally flat, so $N'=N$ then. Consequently,
the IP leads to deviations from Einstein gravity only if $N'\ne N$ locally.
From strict analogy with the Einstein equation, in which curvature is linearly
proportional to any energy-momentum, all information on energy-momentum seems
equally relevant to the imposition of gravity. Indeed, the mass of a particle
matters just as much as any of its four space-time coordinates to the force of
gravity. Under this democratic approach, the IP must be carried by $N'_O/N_O$,
for an observer $O$ that experiences the gravitational force of some system, as
follows.

Einstein gravity, for a fixed gravitational constant $G$, is given by
$${\bf G}_{\mu\nu}/8\pi G {\bf T}_{\mu\nu}=1,$$
in the usual notation. Under the IP,
$$[{\bf G}_{\mu\nu}/8\pi G {\bf T}_{\mu\nu}]_O=N'_O/N_O.$$
Information gravity (IG) seems an appropriate name for this conjectured form of
gravity.
No free parameters are introduced, $N'_O/N_O$ adjusts the strength of gravity,
and general covariance is preserved.
It is fascinating that $N'_O/N_O$ is a rational number and is related to a
local geometric quantity\cite{3}.

\section{Tests of Information Gravity}

In general, the value of $N'_O/N_O$ requires thought, but simple cases exist.
The $n_*\ge 10^{10}$ stars inside a typical galaxy form a mostly collisionless
dynamic system. One has $N'_O/N_O\approx w'n_*/wn_*$ and $N'_O/N_O\approx 6$:
For an observer in orbit around a galaxy one requires a mass, a spin and four
space-time coordinates per star, $w'=6$ and $x'=0$.
From a non-gravitational perspective one merely needs to assign a mass to each
star, $w=1$, and provide a global characterization of a close to collisionless
system through the mean free path and mean energy, so $x=2<<n_*$.
In this, the internal structure of individual stars is negligible (order
$1/n_*$), and collisional interstellar matter (with $w'=w=7$) is ignored.
Overall, IG yields 6 times stronger accelerations, pertinent to dark matter
theories.

The Earth moves around the Sun, which yields $N'_O/N_O=1$ and no deviation from
Einstein gravity: One assigns a gravitational mass, charge, spin and four
space-time coordinates to every elementary particle that is contained by the
Sun and adds to its gravitational pull, $w'=7$ and $x'=0$.
Non-gravitational interactions inside the Sun are dominated by small scale
thermal collisions and these require an inertial mass, charge, spin and four
space-time coordinates per elementary particle, so $w=w'$ and $x=x'$. A small
enhancement in gravitational acceleration, of order
$(n_{\rm Venus}+n_{\rm Mercury})/n_{\rm Sun}\sim 10^{-6}$, is expected due to
Venus and Mercury that move collisionlessly within the Earth's orbit. Also,
$N'_O/N_O<1$ would be reminiscent of a cosmological constant. Finally, the IP
should be applicable to quantum information and to all fundamental forces.
E.g., an information adjusted $G'\equiv GN'_O/N_O$ could be implemented for
other coupling constants as well: ``Information governs forces''.

%{\small

\end{document}